\documentclass{article}
\usepackage{emulateapj}
\usepackage{epsfig}

\slugcomment{Accepted version -- 2000 May 3}

\begin{document}

\title{The Detection of Wind Variability in Magellanic Cloud O Stars}

\author{D.\ Massa\altaffilmark{1}, A.\ W.\ Fullerton\altaffilmark{2}
\altaffilmark{3}, J.\ B.\ Hutchings\altaffilmark{4}, D.\ C.\ 
Morton\altaffilmark{4}, G.\ Sonneborn\altaffilmark{5}, A.\ J.\ 
Willis\altaffilmark{6}, L.\ Bianchi\altaffilmark{3}, K.\ R.\ 
Brownsberger\altaffilmark{7}, P.\ A.\ Crowther\altaffilmark{6}, T.\ P.\ 
Snow\altaffilmark{7}, D.\ G.\ York\altaffilmark{8}}

\altaffiltext{1}{Raytheon ITSS, Code 681.0, NASA's GSFC, Greenbelt, 
 MD 20771, derck.massa@gsfc.nasa,gov}
\altaffiltext{2}{Dept. of Physics \& Astronomy, University of Victoria, 
 P.O.\ Box 3055, Victoria, BC, V8W 3P6, Canada}
\altaffiltext{3}{Dept. of Physics \& Astronomy, The John Hopkins University, 
 3400 N.\ Charles St., Baltimore, MD 21218}
\altaffiltext{4}{D.A.O., Herzberg Institute of Astrophysics, National 
 Research Council, 5071 West Saanich Rd., Victoria, BC, Canada, V8X 
 4M6}
\altaffiltext{5}{Lab. for Astro. \& Solar Phys., Code 681.0, NASA's GSFC, 
 Greenbelt, MD 20771}
\altaffiltext{6}{Dept.\ of Phys.\ \& Astr., University College London, 
 Gower St., London WC1E 6BT, UK}
\altaffiltext{7}{University of Colorado, CASA-Campus Box 389, Boulder, 
 CO 80309}
\altaffiltext{8}{Univ. of Chicago, Astronomy \& Astrophysics Center, 5640 
 S. Ellis Ave., Chicago, IL 60637}

\begin{abstract}
We present Far Ultraviolet Explorer ({\it FUSE}) spectra for three 
Magellanic Cloud O stars (Sk 80, Sk $-67^\circ$05 and Sk $-67^\circ$111) 
with repeated observations. The data demonstrate the capabilities of 
{\it FUSE}\/ to perform time-resolved spectroscopy on extragalactic stars.  
The wavelength coverage of {\it FUSE}\/ provides access to resonance lines 
due to less abundant species, such as sulfur, which are {\em unsaturated} in 
O supergiants.  This allows us to examine wind variability at all velocities 
in resonance lines for stars with higher mass loss rates than can be studied 
at longer ($\lambda \geq 1150$\AA) wavelengths.  The {\it FUSE}\/ wavelength 
range also includes resonance lines from ions which bracket the expected 
dominant ionization stage of the wind.  Our observations span 1-4 months 
with several densely sampled intervals of 10 hours or more.  These 
observations reveal wind variability in all of the program stars and 
distinctive differences in the ionization structure and time scales of the 
variability.  Sk $-67^\circ$111 demonstrates significant wind variability on 
a time scale less than 10 hours and the coolest O star (Sk $-67^\circ$05) 
exhibits the largest variations in O {\sc vi}.  
\end{abstract}


\section{Introduction}

Not long after the first observations of UV wind lines by Morton (1967), 
repeated observations with the {\it Copernicus}\/ satellite discovered that 
these lines were time variable (York et al.\ 1977, Snow 1977).  Subsequent 
{\it IUE}\/ observations established that discrete absorption components 
(DACs) are ubiquitous in hot stars with well-developed, but unsaturated 
wind lines (e.g., Howarth \& Prinja 1989) and that the presence of DACs is 
indicative of wind variaility.  Later work by Kaper et al.\ (1996) 
demonstrated the universality UV wind line variability directly, through 
time series observations of selected O stars.  Massa et al.\ (1995) and 
Kaper et al.\ determined that the wind lines in many stars vary on their 
stellar roation time scale, verifying Prinja's (1988) suggestion that 
wind line activity is linked to the stellar rotation period.  This has 
lead researchers to consider physical explanations which rely on phenomena 
present on the stellar surface, such as magnetic fields (Henrichs et al.\ 
1998) or non-radial pulsations (Gies et al.\ 1999).  Whatever the origin, 
Owocki et al.\ (1995) have shown that winds with large scale spiral patterns 
are consistent with some of the observations (Fullerton et al.\ 1997, Kaper 
et al.\ 1999), although other aspects of the variability remain unexplained. 

All of the previous results were deduced from observations of Galactic 
stars.  Although there is no reason to doubt similar variability in normal 
extragalactic stars (Prinja \& Crowther 1998 have detected DACs in several 
Magellanic Cloud stars), direct observations of wind variability have not 
been obtained.  Beside demonstrating its universality, observing wind 
variability in Large and Small Magellanic Cloud (LMC, SMC) stars with {\it 
FUSE}\/ is important in other respects.  First, the lower metallicity of the 
clouds provides an opportunity to study how instabilities in line driven 
winds are affected by abundances.  Second, the ions available to {\it 
FUSE}\/ complement those normally observed at longer wavelengths.  This is 
especially true for the S {\sc iv} and S {\sc vi} lines which can be used to 
determine the ionization structure of wind variability.  Third, because 
sulfur is less abundant than ions with comparable ionization potentials 
accessible at longer wavelengths, its resonance lines are less optically 
thick and can be used to probe activity at deeper levels in the winds.

\medskip
\begin{tabular}[*t]{|l|l|c|c|} 
\multicolumn{4}{c} {{\bf Table 1}} \\ \hline
\multicolumn{4}{|c|} {\bf {Program Stars}} \\ \hline
    Name	   & Sp Ty$^a$	  & $v_{\infty}$ (km s$^{-1}$) & $V$ (mag) \\ \hline
 Sk 80		   & O7 Iaf+	  & 1400$^b$  & 12.36 \\
 Sk $-67^\circ$05  & O9.7 Ib	  & 1665$^c$  & 11.34 \\
 Sk $-67^\circ$111 & O7 Ib(f)	  & 1800$^b$  & 12.57 \\ \hline
\end{tabular} \\
\small
\noindent $^a$ Spectral types from Fitzpatrick (1988) and Walborn (1977) \\
\noindent $^b$ Bianchi et al.\ (2000) \\
\noindent $^c$ Patriarchi \& Perinotto (1992)
\normalsize
\medskip

\section{Observations}

As part of in-orbit checkout (IOC), extended observational sequences were 
performed on three O stars in the Magellanic Clouds (Table 1).  The 
observations were obtained during several visits over intervals spanning 1 
to 4 months.  They include a few continuous time streams of several hours 
and several cases with multiple exposures obtained over the course of a 
single day.  

All of the observations were obtained in time tagged mode through the low 
resolution apertures (Moos et al.\ 2000).  Because the observations were 
obtained early in the mission, the SiC channels were not aligned.  
Consequently, we only have repeated observations for the LiF channels and 
currently lack repeated observations for the important S {\sc vi} $\lambda 
\lambda 940$ doublet.

Because much of the IOC data were obtained to determine the locations of the 
apertures for the LiF and SiC channels, the position of the target was moved 
in the aperture during an observation.  In fact, the target was often 
completely out of the LiF aperture for part of the observation.  Due to 
these intentional, large image motions, normal pipeline reduction was not 
feasible.  Instead, the data were reduced by a suite of IDL procedures which 
proceeded in two stages.  In the first stage, the photon lists were divided 
into five minute sub-exposures.  A spectrum was extracted for each 
sub-exposure by summing all counts at a constant $x$ (where $x$ is in the 
dispersion direction) within a pre-specified $y$ range, to obtain the total 
counts at each $x$.  Background corrections are insignificant for our bright 
program stars and were neglected.  During this stage, the total counts in a 
pre-specified wavelength band were determined for each sub-exposure and a 
``postage-stamp'' 64$\times$64 binned image of the detector was produced.  
In the second processing stage, the individual sub-exposures were examined 
for event bursts (see Sahnow et al.\ 2000) and uniformity of flux levels.  
Sub-exposures that were affected by bursts or which had discrepant flux 
levels were discarded.  Finally, the remaining sub-exposures were aligned by 
using their mean as a template and then shifting and fitting (in a least 
squares sense) the individual sub-exposures to the mean.  The minimum 
$\chi^2$ shifts were applied to each sub-exposure to create the set of 
aligned spectra.

\section{Results}

Since we only have repeated observations for the LiF channels, we 
concentrate on the LiF1A data ($987 \leq \lambda \leq 1082$\AA) which 
contain the S {\sc iv} and O {\sc vi} wind lines. The data are displayed in 
Figures 1--3 as {\it dynamic spectra}.  These are 2-dimensional images of 
time ordered spectra.  The ordinates are sequential spectrum number and the 
abscissae are velocity relative to the blue component of the doublet 
(adjusted for the stellar radial velocity).  Each horizontal strip is a 5 
minute sub-exposure normalized by the {\it sample mean spectrum} (shown at 
the bottom).  Sequential stacking avoids the large time gaps which appear 
when irregularly sampled data are displayed on a linear time scale.  Each 
figure also contains a {\it temporal plot}\/ which gives the relative 
observation time for each sub-exposure.  A nearly vertical line implies that 
several spectra were obtained over a very short time period.  

\medskip
\noindent {\bf Sk 80 = AV 232} (Figure 1): This SMC star has distinctively 
sub-solar metallicity (see Fullerton et al.\ 2000).  We obtained 4 sets of 
observations between 1999 September 25 and 1999 November 17.  Notice that 
although the S {\sc iv} $\lambda 1063$ line and $\lambda \lambda 1073$ 
doublet are stable throughout the individual observations (which span up to 
12 hours) they vary strongly from one set of observations to the next.  
There is also some indication of an overall brightening of the object at 
$\Delta t = 30$ days as well as marginal evidence for a change in the O 
{\sc vi} $\lambda \lambda 1030$ doublet (which is intrinsically very weak).  
However, given the nature of the observational sequence, absolute flux 
levels are uncertain.

The form of the variability affects the entire profile, but the current 
data set is not suitable for addressing the temporal evolution of wind 
variability because adjacent observations are separated by several wind 
flushing times.  The presence several interstellar molecular hydrogen lines 
throughout the region and strong Ly $\beta$ absorption affecting the blue 
component of O {\sc vi} (see Fullerton et al.\ 2000), make it difficult to 
verify whether an isolated absorption observed in one component of a doublet 
is part of a DAC.

\medskip
\noindent {\bf Sk $-67^\circ$111} (Figure 2): This is an LMC analog of Sk 
80 which has a higher metallicity (Fullerton et al.\ 2000).  We obtained 
spectra on 1999 September 26 and 1999 October 31, with the September set 
spanning 10 hours. In this case, the overall S {\sc iv} variability is 
weaker than in Sk 80, but there is better evidence for O {\sc vi} 
variability.  However, it is especially interesting that the S {\sc iv} 
lines clearly vary during the 10 hour Sept.\ observation.  Throughout this 
time, a broad absorption band centered near $-1200$ km s$^{-1}$ in S {\sc 
iv} weakened, narrowed and propagated toward higher velocity.  A similar 
feature is discernible in the red component of O {\sc vi}, but the blue 
component is lost in Ly $\beta$ absorption.

\medskip
\noindent {\bf Sk $-67^\circ$05 = HD 268605} (Figure 3): This LMC star is 
more than 2 full spectral sub-classes later than the others (Table 1).  Our 
data span nearly 4 months, from 1999 August 23 to 1999 December 13.  In this 
case, the O {\sc vi} variability is quite strong, and in phase with much 
weaker S {\sc iv} variability.  There is no evidence for short term 
variability. 

\section{\bf Discussion}

Although we lack well-sampled temporal coverage and repeated observations 
for S {\sc vi} at this time, the current data still provide a glimpse of the 
capabilities of {\it FUSE}\/ to perform wind variability studies.  The 
combination of {\it FUSE}\/ wavelength coverage and the reduced metallicity 
of the Magellanic Cloud stars enabled us to examine wind variability in new 
density regimes and new ionization states.  In many respects, S {\sc iv} and 
Si {\sc iv} $\lambda \lambda$ 1400 are sensitive to similar plasmas.  
However, the reduced abundance of sulfur allows us to probe more massive 
flows.  Both O {\sc vi} and N {\sc v} sample very hot plasma.  However, the 
wider separation of the O {\sc vi} $\lambda \lambda 1032, 1038$ doublet 
removes the complications introduced by the overlap of the N {\sc v} 
$\lambda \lambda 1240$ components in many stars (however, interstellar Ly 
$\beta$ contaminates O {\sc vi} $\lambda 1032$).  These features allowed us 
to detect wind variability at intermediate velocities in O7 supergiants.  In 
Galactic O7 supergiants, all of the wind lines in the {\it IUE}\/ or {\it 
HST}\/ range are saturated, and it is not until O6 and earlier supergiants 
that Si {\sc iv} desaturates enough for variability to be detected (see 
$\lambda$ Cep in Kaper et al., 1996).  Even then, N {\sc v} remains 
completely saturated, so ionization information cannot be obtained.  In 
contrast, the P~Cygni profiles of S {\sc iv} are well developed and 
unsaturated in the {\it FUSE}\/ spectra of both O7 supergiants. O {\sc vi} 
is similarly well-suited studying variations in spectra of Sk 
$-67^\circ$111, but is too weak to be useful in Sk 80.

Several aspects of the current data are noteworthy.  First, to provide an 
indication of the magnitude of the physical changes implied by the 
morphological variability, consider the results from Bianchi et al.\ (2000) 
for S {\sc iv} in Sk 80 and Sk $-67^\circ$111.  Analyzing the same 
observations displayed here, they determined that the line of sight wind 
column density varied by 68\% in Sk 80 and 45\% in Sk $-67^\circ$111.  We 
see, therefore, that the variability is not a small perturbation on top of 
an otherwise steady flow.  Instead, {\em variability is a fundamental 
property of the winds}.  Second, the {\em wind lines vary in every 
Magellanic Cloud O star with repeated observations} separated by more than a 
few days, a result which is similar to that observed in Galactic O stars 
(Kaper et al.\ 1996).  Third, although the S {\sc iv} variability in Sk 80 
is stronger than in Sk $-67^\circ$111, its O {\sc vi} variability is weaker, 
if present at all -- an effect that may be due to the relative abundances of 
the two stars.  Fourth, although Sk $-67^\circ$05 is cooler than either Sk 
80 or Sk $-67^\circ$111, its O {\sc vi} variability is much stronger.  This 
implies that the variability we happened to observe is of a higher 
ionization state than in the two O7 stars.  Fifth, while there is no 
evidence of short term ($\leq 1$ day) variability in two of the program 
stars, it is clearly present in Sk $-67^\circ$111 on time scales similar to 
those observed for the Galactic O6 supergiant $\lambda$ Cep (Kaper et al., 
1996).  

From our very limited data, it appears that the overall wind variability of 
the Magellanic Cloud O stars is similar to Galactic stars.  However, longer 
and better sampled time series will be needed to determine whether the time 
scale of the variability is related to the rotation period of the stars, as 
it is in many Galactic O stars and B supergiants.

\acknowledgments
This work is based on data obtained for the Guaranteed Time Team by the
NASA-CNES-CSA {\it FUSE}\/ mission operated by the Johns Hopkins University.
Financial support to U.S.  participants has been provided by NASA contract
NAS5-32985.

\begin{figure}[h]
\centerline{\hbox{
}}
\caption{Dynamic spectrum for Sk 80.  The ordinate is sequential spectrum 
number and the abscissa is wavelength.  Each horizontal strip is a 5 min.\ 
sub-exposure normalized by the sample mean spectrum (shown at the bottom).  
At right, is a temporal plot, which gives the relative time of each 
sub-exposure.  The rest wavelengths of the O {\sc vi} doublet and the S 
{\sc iv} lines are indicated on the mean spectrum.\label{av232}}
\end{figure}
\begin{figure}[h]
\centerline{\hbox{
}}
\caption{Same as Figure 1 for the LMC O7 supergiant Sk $-67^\circ$111.
\label{sk111}}
\end{figure}
\begin{figure}[h]
\centerline{\hbox{
}}
\caption{Same as Figure 1 for the LMC O9.7 supergiant Sk $-67^\circ$05.
\label{sk05}}
\end{figure}

\end{document}